\documentclass[doublecol]{epl2} 

\usepackage{graphicx}
\usepackage{float}
\usepackage{amsmath, amssymb}
\usepackage{lmodern}
\usepackage[T1]{fontenc}
\usepackage[utf8]{inputenc}
\usepackage[english]{babel}
\usepackage{color}

\def\lsim{\mathrel{\rlap{\lower4pt\hbox{\hskip1pt$\sim$}}
    \raise1pt\hbox{$<$}}}                
\def\gsim{\mathrel{\rlap{\lower4pt\hbox{\hskip1pt$\sim$}}
    \raise1pt\hbox{$>$}}}                

\title{Death by starvation in May-Leonard models}
\shorttitle{Death by starvation in May-Leonard models} 

\author{P.P. Avelino\inst{1,2} \and B.F. de Oliveira\inst{3}}
\shortauthor{P.P. Avelino and B.F. de Oliveira}

\institute{                    
  \inst{1} Instituto de Astrof\'{\i}sica e Ci\^encias do Espa{\c c}o, Universidade do Porto, CAUP, Rua das Estrelas, PT4150-762 Porto, Portugal \\
  \inst{2} Departamento de F\'{\i}sica e Astronomia, Faculdade de Ci\^encias, Universidade do Porto, Rua do Campo Alegre 687, PT4169-007 Porto, Portugal \\
  \inst{3} Departamento de F\'{\i}sica, Universidade Estadual de Maring\'a, Av. Colombo 5790, 87020-900 Maring\'a, PR, Brazil
}
  
\pacs{87.23.Cc}{Population dynamics and ecological pattern formation}
\pacs{87.23.Kg}{Dynamics of evolution}
\pacs{87.23.-n}{Ecology and evolution}

\abstract{
We consider the dynamics of spatial stochastic May-Leonard models with mutual predation interactions of equal strength between any two individuals of different species. Using two-dimensional simulations, with two and three species, we investigate the dynamical impact of the death of individuals after a given threshold number of successive unsuccessful predation attempts. We find that the death of these individuals can have a strong impact on the dynamics of population networks and provide a crucial contribution to the preservation of coexistence.}

\begin{document}

\maketitle

\section{Introduction}

Competition is ubiquitous in nature, playing a fundamental role on the regulation of biodiversity.  It is also a major driving force behind evolutionary change through natural selection. The simplest competition models, inspired in the pioneering work by Lotka and Volterra, and May and Leonard \cite{1920PNAS....6..410L,1926Natur.118..558V,May-Leonard}, consider the dynamics of two or three species subject to interspecific predation (or selection), mobility and reproduction interactions. Despite their simplicity, these models (see \cite{2014-Szolnoki-JRSI-11-0735,2018JPhA...51f3001D} for recent reviews) incorporate some of the main ingredients responsible for the observed dynamics of many biological systems, and are able to reproduce some the dynamical features of specific biological systems with a limited number of species  \cite{lizards,2002-Kerr-N-418-171,bacteria,Reichenbach-N-448-1046}.

More complex competition models, involving more species \cite{Szabo2008,Hawick2011,Hawick_2011,Avelino-PRE-86-031119,Avelino-PRE-86-036112} and/or additional interactions \cite{Peltomaki2008,2010-Wang-PRE-81-046113,2014-Cianci-PA-410-66,2010-Yang-C-20-2,2017-Souza-Filho-PRE-95-062411}, have also been investigated in recent years, revealing a plethora of complex dynamical spatial structures \cite{2010-Ni-PRE-82-066211,Avelino-PLA-378-393,Avelino-PRE-89-042710,2017-Avelino-PLA-381-1014,2018-Avelino-EPL-121-48003}, diverse scaling regimes \cite{Avelino-PRE-86-036112,PhysRevE.96.012147} and phase transitions \cite{PhysRevE.76.051921, 2001-Szabo-PRE-63-061904, 2004-Szabo-PRE-69-031911, 2004-Szolnoki-PRE-70-037102, 2007-Perc-PRE-75-052102, 2008-Szabo-PRE-77-011906, 2011-Szolnoki-PRE-84-046106, 2013-Vukov-PRE-88-022123, 2018-Bazeia-EPL-124-68001}. In some of these competition models species coexistence may last for an arbitrary amount of time, while in others it is transient. Coexistence-promoting mechanisms, responsible for maintaining coexistence over long periods of time, are usually associated to the ability of the species to increase (decrease) their population in response to negative (positive) perturbations to their typical abundances \cite{siepielski_mcpeek_2010}. Among these, density-dependent mortality \cite{gross_edwards_feudel_2009,metz_sousa_valencia_2010} has been claimed to have a positive impact in promoting species coexistence (see also \cite{holt_1985,10.1086/323113,mittelbach_hall_dorn_garcia_steiner_wojdak_2004} for a discussion of the impact of density-independent mortality).

Here, we investigate a sub-class of spatial stochastic May-Leonard models characterized by mutual predation interactions of equal strength between any two individuals of different species. In their standard version, the dynamics of these models results in a network of one-species domains whose dynamics is curvature driven, with the characteristic size of the network of one-species domains growing proportionally to $t^{1/2}$ \cite{Avelino-PRE-86-031119} --- $t$ being the physical time. However, in practice, this growth is limited by the size of the simulation boxes, thus resulting in a limited period of coexistence. 

The main aim of this letter is the determination of the impact on population dynamics of the death by starvation of individuals after a given number of successive unsuccessful predation attempts. We shall start by introducing our set of models in the following section. We then study the impact of death by starvation on the dynamics of initially flat and circular interfaces between spatial domains occupied by individuals of competing species, as well as the two-dimensional dynamics of population networks starting from random initial conditions. We will show that, under certain conditions, death by starvation  prevents the endless growth of the characteristic length scale of the network of one-species spatial  domains, acting as a coexistence-promoting mechanism.

\section{Models \label{sec2}}

In this letter we shall investigate the dynamics of May-Leonard models with mutual predation interactions of equal strength between any two individuals of different species. To this end, we shall perform square lattice simulations with periodic boundary conditions in which each one of its ${\mathcal N}$ sites may be either empty or occupied by a single individual. The species are labelled by the number $i$ (or $j$), with $i,j=1,...,N$ --- in this letter we shall only consider models with two or three species ($N=2$ or $N=3$). Empty sites shall be denoted by $\otimes$. The number of individuals of the species $i$ and the number of empty sites will be denoted by $I_i$ and $I_\otimes$, respectively --- the density of individuals of the species $i$ and the density of empty sites shall be defined by $\rho_i=I_i/{\mathcal N}$ and $\rho_{\otimes} = I_\otimes/{\mathcal N}$, respectively. The possible  interactions are predation
\begin{equation}
i\ j \to i\ \otimes\,, \nonumber
\end{equation}
mobility 
\begin{equation}
 i\ \odot \to \odot\ i\,, \nonumber
\end{equation}
and reproduction 
\begin{equation}
 i\ \otimes \to ii\,, \nonumber
\end{equation}
where $\odot$ represents either an individual of any species or an empty space. 

Mobility, reproduction and predation interactions occur with probabilities $m$, $r$ and $p$, respectively (the same for all species). For the sake of definiteness, we shall take $m=0.5$ and $m+p+r=1$ in all the simulations. At each simulation step, the algorithm randomly picks an occupied site to be the active one, randomly selects one of its four adjacent neighbour sites to be the passive one, and randomly chooses an interaction to be executed by the individual at the active position. These three steps are repeated until a possible interaction is selected. If predation is selected, the impossibility of executing an interaction happens when the passive is an empty site or the passive and active individuals are of the same species. On the other hand, if reproduction is selected, the interaction only takes place if the passive is an empty site. Whenever the selected interaction is not executed, the active individual is said to have carried out an unsuccessful interaction  attempt. A generation time (our time unit) is defined as the time necessary for $\mathcal{N}$ successive (and successful) interactions to be completed. 

The non-standard ingredient in our simulations is the death of an individual after a given number ${\mathcal N}_u$ of successive unsuccessful predation attempts --- in our model the most recent number of successive unsuccessful predation attempts of its progenitor is passed on to every newborn individual at the time of birth. This means that the ability of a newborn to survive unsuccessful predation attempts is strongly dependent on the strength of its progenitor at the time of birth (the strength being defined as ${\mathcal N}_u$ minus the latest number of unsuccessful predation attempts).

\begin{figure}[!htb]
	\centering
	\includegraphics{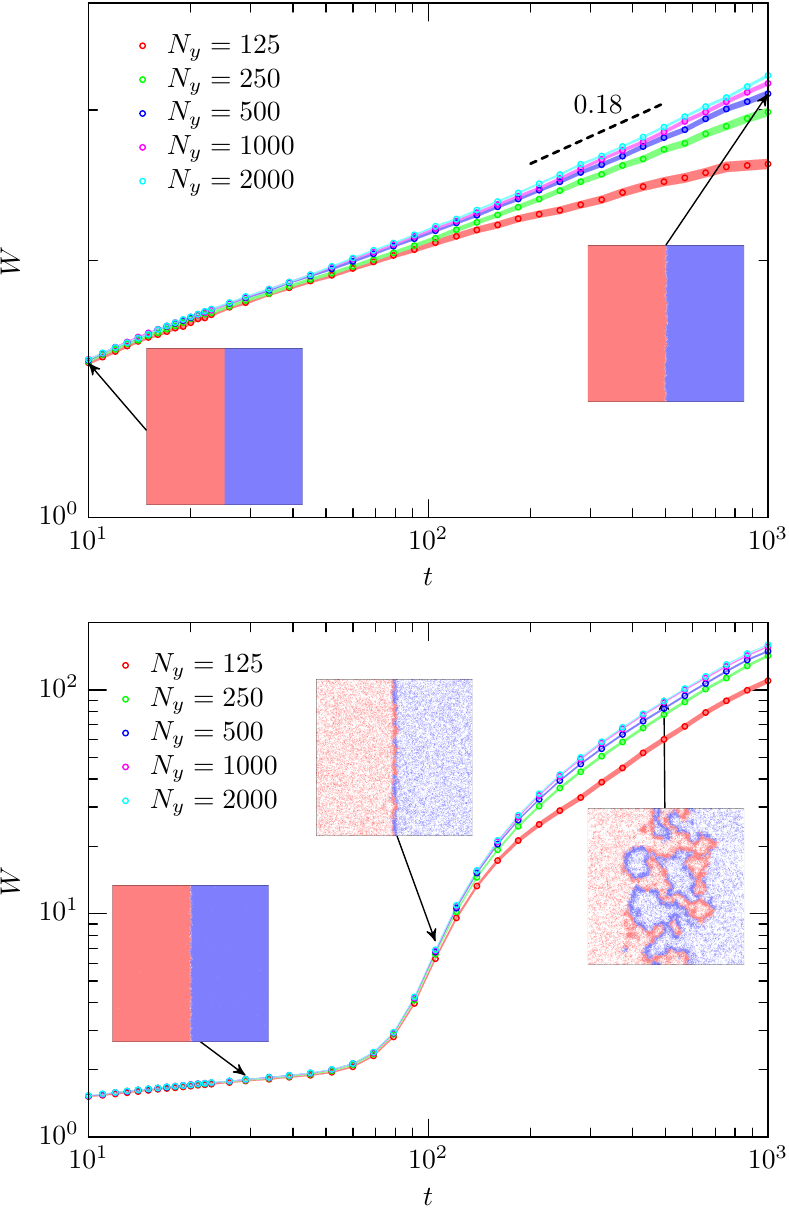}
	\caption{Evolution of the width $W$ of an initially flat interface obtained from $1000$ realizations considering ${\mathcal N}_u=\infty$ (upper panel) and ${\mathcal N}_u=25$ (bottom panel). Notice that the much more significant growth of the width of the interfaces for ${\mathcal N}_u=25$ compared to the standard ${\mathcal N}_u=\infty$ case may also be confirmed in the snapshots, obtained for single realizations of ${\mathcal N}_u=\infty$ and ${\mathcal N}_u=25$ models, shown in the inset panels}
	\label{fig1}
\end{figure}

\section{Dynamics of initially flat/circular interfaces \label{sec3}}

In this section we study the impact that a finite value of ${\mathcal N}_u$ has on the dynamics of spatial stochastic two-species May-Leonard models with mutual interspecific predation interactions of equal strength. To this end, a large number of spatial stochastic numerical simulations has been performed with the following parameters: $m=0.5$, $r=0.3$, $p=0.2$, and ${\mathcal N}_u=25$ (any individual dies after 25 successive unsuccessful predation attempts) or ${\mathcal N}_u=\infty$ (standard case, no deaths by starvation). In this section we shall consider the dynamics of initially flat and circular interfaces, before studying the dynamics of population networks with random initial conditions in the following section.

\begin{figure}[!htb]
	\centering
	\includegraphics{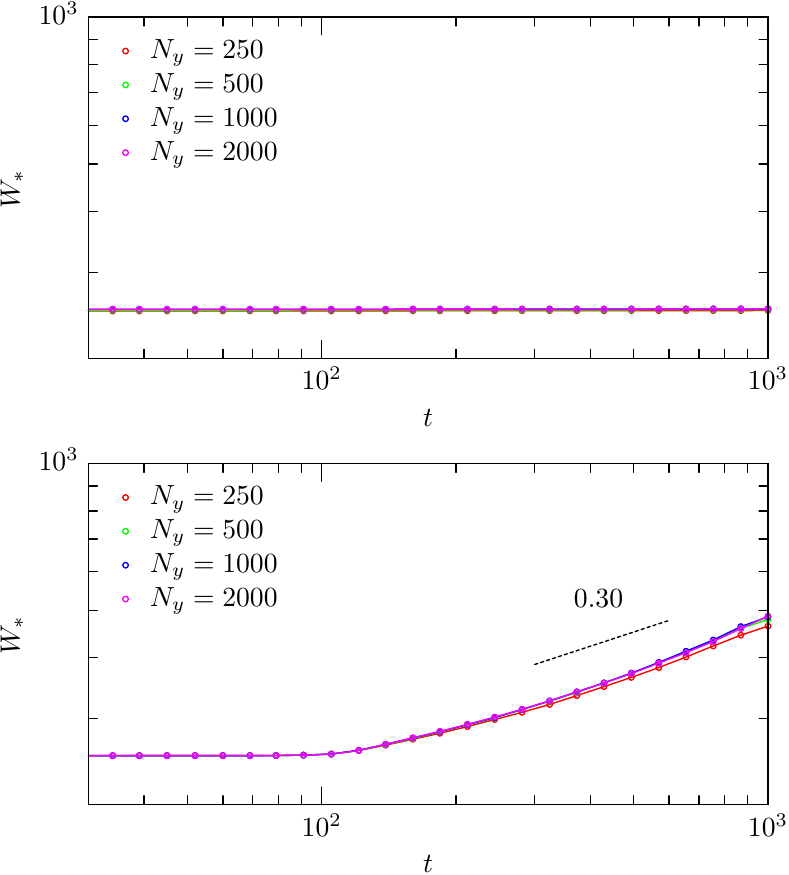}
	\caption{Evolution of the width $W_*$ of an initially flat interface obtained from $1000$ realizations considering ${\mathcal N}_u=\infty$ (upper panel) and ${\mathcal N}_u=25$ (bottom panel). Notice that not only $W_*$ is much less sensitive to the box size than $W$, but also that the growth of $W_*$ with time observed in the case with ${\mathcal N}_u=25$ is absent for ${\mathcal N}_u=\infty$.}
	\label{fig2}
\end{figure}

\subsection{Initially flat interface}

Here we consider the dynamics of an initially flat interface separating the left and right halves of the lattice which are initially fully occupied by individuals of the red (1) and blue (2) species, respectively.  Our simulations are performed on a $N_x \times N_y$ lattice. The position ($k_W(l)$) of the interface for each row $l$ may be found as the value of $k_W(l)$ that minimizes the sum
\begin{equation}
F(k_W[l]) = \sum_{k=1}^{N_x} \left(S_{kl}-ST[k-k_W]\right)^2\,
\end{equation}
for each value of $l$. Here, $ST[x]$ is the step function, defined as $ST=1$ for $x<0$, $ST=0$ for $x=0$, and $ST=-1$ for $x>0$, and $S_{kl}=0$, $S_{kl} = +1$, or $S_{kl} = -1$ depending on whether the site of coordinates $(k,l)$ is empty, occupied by an individual of the red (1) species, or occupied by an individual of the blue (2) species. We shall follow refs. \cite{2017-Brown-PRE-96-012147, 2013-Roman-PRE-87-032148, 2012-Roman-JSMTE-2012-p07014}, and define the interface width as
\begin{equation}
W(t)=\sqrt{\frac{1}{N_y} \sum_{l=1}^{N_y} \left(k_W[l] - \langle k_W \rangle\right)^2}\,,
\end{equation}
where
\begin{equation}
\langle k_W \rangle = \frac{1}{N_y} \sum_{l=1}^{N_y} k_W[l] \,.
\end{equation}

\begin{figure}[!htb]
	\centering
	\includegraphics{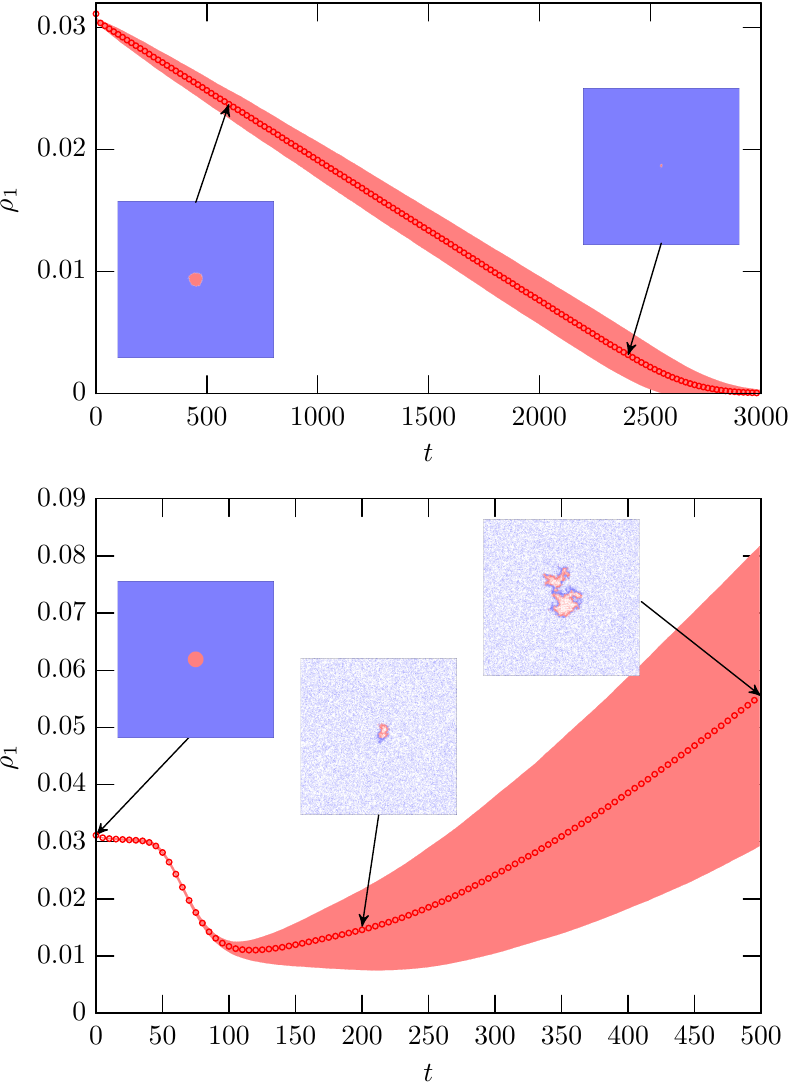}
	\caption{Evolution of the average density $\rho_1$ of individuals of the red species ($1$) obtained using $1000$ realizations of the evolution of an initially circular domain containing individuals from that species surrounded by an outer domain containing individuals of the blue species ($2$), considering ${\mathcal N}_u= \infty$ (upper panel) and ${\mathcal N}_u=25$ (bottom panel). The solid red line shows the average value of $\rho_1$ as a function of the number of generations $t$ while the shaded region represents the sample standard deviation. Notice that the collapse of the circular domain, which takes place for ${\mathcal N}_u=\infty$, is not observed for ${\mathcal N}_u=25$. Instead, the inset panels displaying snapshots of a single realization taken at three different times show that for ${\mathcal N}_u=25$ there is a significant departure of circular symmetry and that, after an initial shrinking stage, the domain may grow and split into separate subdomains.}
	\label{fig3}
\end{figure}

Figure \ref{fig1} illustrates  the evolution of the width of an initially flat interface obtained from $1000$ realizations considering ${\mathcal N}_u=\infty$ (upper panel) and ${\mathcal N}_u=25$ (bottom panel).  Notice that, after an initial transient stage, the growth of the width of the interfaces for ${\mathcal N}_u=25$ is much faster compared to that of the standard ${\mathcal N}_u=\infty$ case --- the evolution approaching a scaling regime with $W \propto t^{0.18}$ at large $t$ for ${\mathcal N}_u=\infty$. This much faster growth may also be confirmed in the snapshots, obtained for single realizations of the  ${\mathcal N}_u=\infty$ and ${\mathcal N}_u=25$ models, shown in the inset panels. The inset panels show the development of complex dynamical structures along the interfaces for ${\mathcal N}_u=25$, leading to rougher interfaces compared to the standard ${\mathcal N}_u=\infty$ case. It is also clear from the inset panels that for ${\mathcal N}_u=25$ death by starvation results in a relatively low constant average density of individuals away from the interfaces --- the average density being reached when the equilibrium between the mortality and reproduction rates is attained. This is the key property of models with finite ${\mathcal N}_u$, which is responsible for the growth of the interface thickness and roughness with time observed in Fig. \ref{fig1}. 

Figure \ref{fig2} shows the average evolution of the width of an initially flat interface with time obtained from $1000$ realizations with ${\mathcal N}_u=\infty$ (upper panel) and ${\mathcal N}_u=25$ (bottom panel), considering an alternative definition of interface width. In this case, the interface width $W_*$ defines the interval of $k$ for which the abundance of the two species is bellow $68\%$. Fig. \ref{fig2} shows that, not only $W_*$ is much less sensitive to the box size than $W$, but also that the growth of $W_*$ with time observed in the ${\mathcal N}_u=25$ case is absent for ${\mathcal N}_u=\infty$. For ${\mathcal N}_u=25$ the evolution approaches a  scaling regime with $W_* \propto t^{0.3}$ at large $t$.

Although the two definitions of interface width are physically distinct --- $W_*$ being much less sensitive than $W$ to small wavelength fluctuations which do not introduce large modifications to the interface profile at each row --- both evolve very differently in the ${\mathcal N}_u=25$ and ${\mathcal N}_u=\infty$ cases, thus capturing the impact of death by starvation on interface dynamics.

\subsection{Initially circular interface}

Let us now consider the dynamics of an initially circular interface. Figure \ref{fig3} illustrates the evolution of the average density $\rho_1$ of individuals of the the red species ($1$) obtained using $1000$ realizations of the evolution of an initially circular domain containing individuals of that species surrounded by an outer domain containing individuals of the blue species ($2$), considering ${\mathcal N}_u=\infty$ (upper panel) and  ${\mathcal N}_u=25$ (bottom panel). The solid red line shows the average value of $\rho_1$ as a function of the number of generations $t$ while the shaded region represents the sample standard deviation. 

The upper panel of Fig.  \ref{fig3} shows that if  ${\mathcal N}_u = \infty$ the initially circular domain always collapses. In this case, a standard curvature dominated regime is recovered, with the domain wall area being roughly proportional to $t-t_c$, for $t \le t_c$ ($t_c$ being the time of collapse). This explains the approximately linear dependence of the average density of the inner species (1) on time.

The bottom panel of Fig. \ref{fig3} shows that, contrary to what happens for ${\mathcal N}_u = \infty$, in the ${\mathcal N}_u = 25$ case initially circular domains do not collapse despite the existence of an initial shrinking stage. Instead, the inset panels, displaying snapshots of a single realization taken at three different times, show that for ${\mathcal N}_u=25$ there is a significant departure of circular symmetry and that, after an initial shrinking stage, the initially circular domain may grow and split into separate subdomains.

\begin{figure}[!htb]
	\centering
	\includegraphics{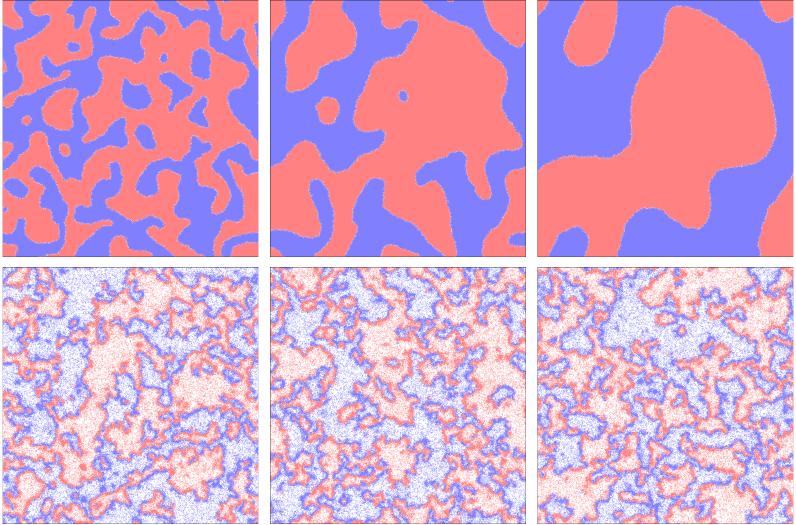}
	\caption{Evolution of the population network in the two species model for ${\mathcal N}_u=\infty$ (upper panels) and ${\mathcal N}_u=25$ (bottom panels). The snapshots of a $1000^2$ simulation of the two and three species model were taken after $1000$ (left panel), $4000$ (middle panel), and $16000$ generations (right panel). Notice that the growth of single species domains, which takes place for ${\mathcal N}_u=\infty$, is not observed in the ${\mathcal N}_u=25$ case.}
	\label{fig4}
\end{figure}

\begin{figure}[!htb]
	\centering
	\includegraphics{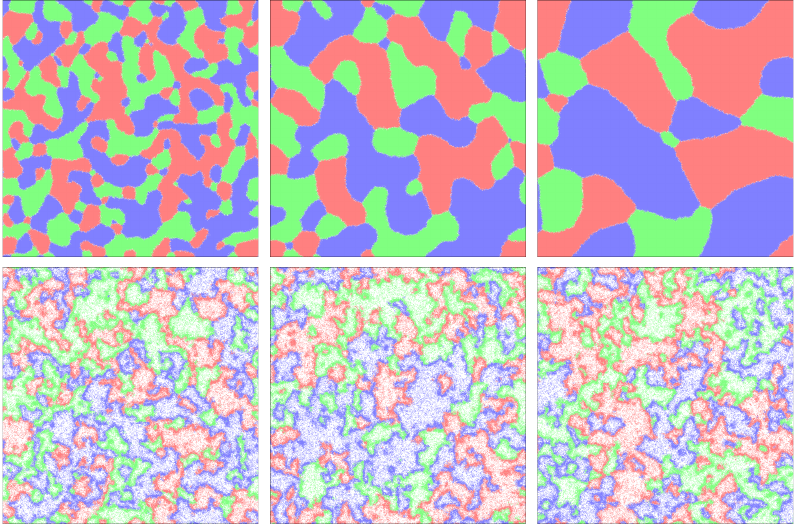}
	\caption{The same as in Fig. \ref{fig4} but for the three species model.}
	\label{fig5}
\end{figure}

\section{Network Simulations \label{sec4}}

In this section we shall consider the results of spatial stochastic numerical simulations with random initial conditions in two spatial dimensions. At the beginning of the simulation each site is either occupied by an individual of any of the $N=2,3$ species or left empty with a uniform discrete probability of $1/(N+1)$  (except if stated otherwise, the following parameter values are assumed: $m=0.5$, $r=0.3$, $p=0.2$). The results of $1000^2$ simulations of the two and three species models for ${\mathcal N}_u=\infty$ (upper panels) and ${\mathcal N}_u=25$ (bottom panels) are illustrated in Figs. \ref{fig4} and \ref{fig5}, respectively. The snapshots were taken after $1000$ (left panels), $4000$ (middle panels), and $16000$ generations (right panels).

\begin{figure}[!htb]
	\centering
	\includegraphics{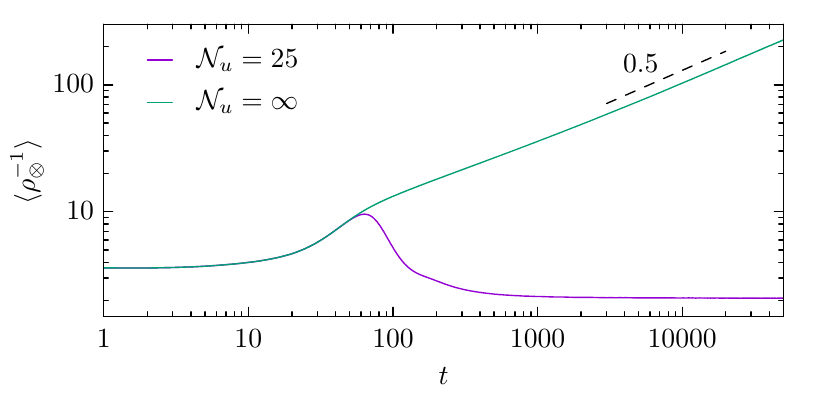}
	\caption{Evolution of $\langle \rho_\otimes^{-1} \rangle$ with time ($\rho_\otimes^{-1}$ being proportional to the characteristic length $L$ of the network). Notice that for ${\mathcal N}_u=\infty$, after an initial transient regime, $\langle \rho_\otimes^{-1} \rangle$ grows proportionally to $t^{1/2}$. On the other hand, for ${\mathcal N}_u=25$ the network attains a regime in which $\langle \rho_\otimes \rangle \sim {\rm const}$, indicating that the average characteristic length scale of the network becomes roughly constant in time. In both cases  the simulations run for $50000$ generations on a $5000^2$ grid. The results were averaged over $25$ simulations.}
	\label{fig6}
\end{figure}

\begin{figure}[!htb]
	\centering
	\includegraphics{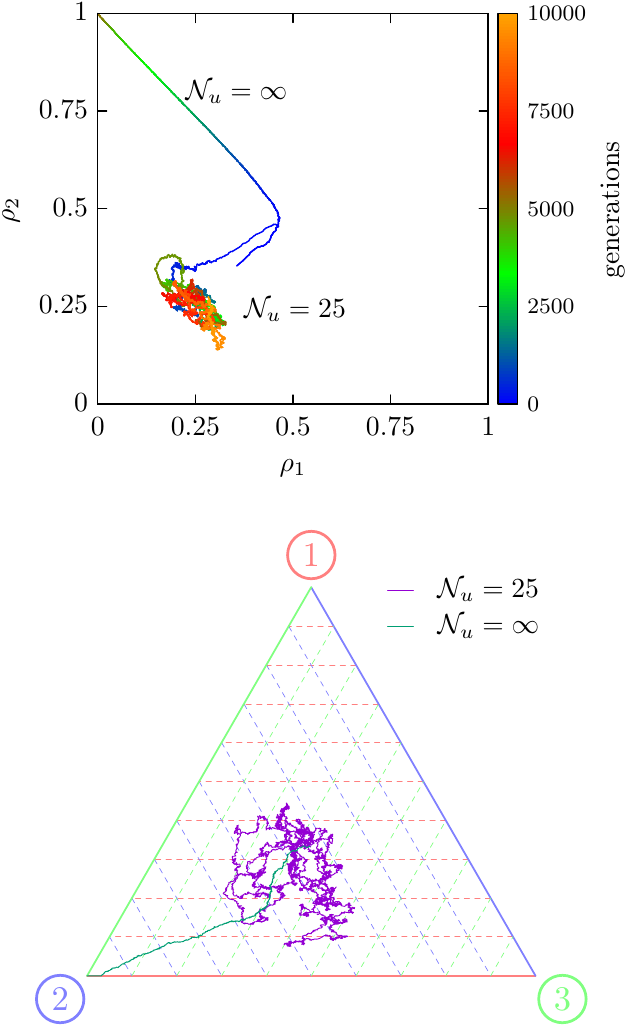}
	\caption{The phase space evolution for a single realization of the standard (${\mathcal N}_u=\infty$) and starvation (${\mathcal N}_u=25$) models with two and three species (top and bottom, panels, respectively panel). Although the initial conditions are the same for both standard and starvation models, long lasting coexistence only occurs in the starvation model. In both case the simulations run for 10000 generations on a $250^2$ grid.}
	\label{fig7}
\end{figure}

In the absence of death by starvation (${\mathcal N}_u=\infty$) there are almost no empty sites deep inside the domains. Empty sites are created only when predation occurs on the interface between competing domains. Empty sites can move around as a result of mobility interactions, but are eventually filled as a result of reproduction. In the ${\mathcal N}_u=\infty$ case, the dynamics is curvature dominated, with the velocity of the interfaces being roughly proportional to their curvature. This has been shown to lead to a population network evolution whose characteristic lengthscale $L$ is roughly proportional to $t^{1/2}$ \cite{Avelino-PRE-86-031119}. This growth of the characteristic scale of the network can be observed in the upper panels of Figs. \ref{fig4} and \ref{fig5}, for simulations with two and three species respectively. Eventually, for ${\mathcal N}_u=\infty$ the size of the domains becomes of the order of the box size and the coexistence is lost (as illustrated in Fig. \ref{fig7} considering a simulation on a smaller grid).

On the other hand, the bottom panels of Figs. \ref{fig4} and \ref{fig5} show that, in the presence of a mortality rate associated to insufficient  predation (${\mathcal N}_u=25$), the average density of individuals has a peak in the interface regions, decreasing towards the interior of the domains, reaching an asymptotic value determined by the equilibrium between death and reproduction. This is responsible for the dynamical behaviour of initially flat and circular interfaces found in the previous sections and is the reason why the characteristic scale of the network does not change significantly from $t=1000$ to $t=16000$, as observed in the bottom panels of Figs. \ref{fig4} and \ref{fig5}.

As discussed before, for ${\mathcal N}_u=\infty$ the empty sites are mainly concentrated on the borders of competing domains. The thickness of the interfaces of empty sites is roughly constant, which implies that the total interface length $L_T$ is roughly proportional to the number of empty spaces $I_\otimes$. On the other hand, the number of domains is roughly proportional the ratio between the total area $A$ and the average domain area $L^2$, where $L$ is the characteristic length scale of the network. It is also proportional to the ratio between the total interface length $L_T$ inside the simulation box and the average domain perimeter (which is proportional to $L$). Therefore, $A/L^2 \propto L_T /L$, thus implying that $L \propto A/L_T \propto {\mathcal N}/I_\otimes =  \rho_\otimes^{-1}$ \cite{Avelino-PRE-86-031119}. Hence, for ${\mathcal N}_u=\infty$ the characteristic length scale of the network is inversely proportional to the number of empty spaces.

Figure \ref{fig6} depicts the evolution of $\langle \rho_\otimes^{-1} \rangle$ (the angle brackets represent an ensemble average) with time for ${\mathcal N}_u=\infty$ (solid green line) and ${\mathcal N}_u=25$ (solid magenta line) --- $\rho_\otimes^{-1}$ being proportional to the characteristic length $L$ of the network.  In both cases the simulations run for $50000$ generations on a $5000^2$ grid. The results were averaged over $25$ simulations. Figure \ref{fig6} shows that in the absence of death by starvation (${\mathcal N}_u=\infty$), and for $t \gsim 50$, the characteristic length scale of the network grows with time as $\langle L \rangle \propto \langle \rho_\otimes^{-1} \rangle \propto t^{1/2}$. On the other hand, in the ${\mathcal N}_u=25$ case the network attains a regime in which $\langle \rho_\otimes \rangle \sim \rm const$, indicating that the average characteristic length scale $\langle L \rangle$ of the network becomes roughly constant in time for $t \gsim 200$.

Figure \ref{fig7} shows the phase space evolution for a single realization of the standard (${\mathcal N}_u=\infty$) and starvation (${\mathcal N}_u=25$) models with two and three species (top and bottom, panels, respectively panel). The initial conditions are the same for both standard and starvation models and, in both cases, the simulations run for 10000 generations in a $250^2$ grid. Figure \ref{fig7} confirms that long lasting coexistence only occurs in the starvation model (${\mathcal N}_u=25$). 

\begin{figure}[!htb]
    \centering
    \includegraphics{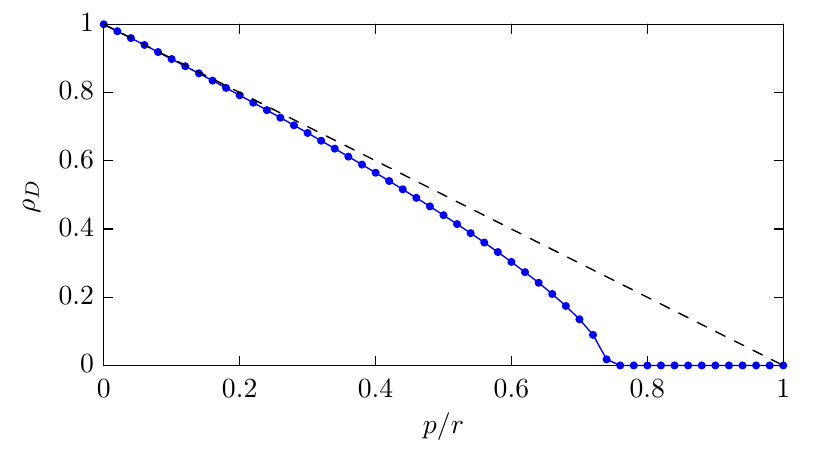}
    \caption{Comparison between the values of the density of individuals $\rho_D$ deep inside the domains obtained using the analytical approximation given by Eq. (\ref{rhod}) (dashed dark line) and those obtained from the numerical results (blue dots connected with a solid line) for various values of $p/r$.}
    \label{fig8}
\end{figure}

\begin{figure}[!htb]
    \centering
    \includegraphics{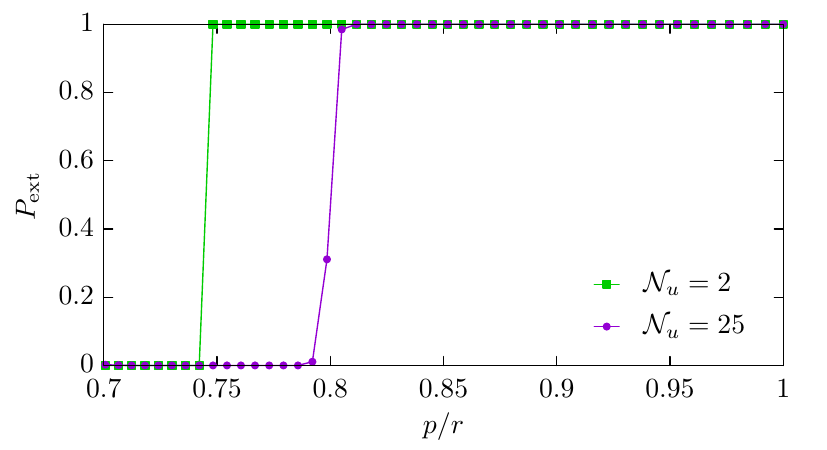}
    \caption{Probability of extinction of at least one species as a function of $p/r$ (assuming that $m=0.5$ and $p+r+m=1$). The results were obtained from an average of $1000$ $200^2$ simulations with two species  ($N=2$) starting from random initial conditions.}
    \label{fig9}
\end{figure}

Far from the borders of sufficiently large spatial domains occupied by a single species individuals die at each predation attempt, independently of the value of ${\mathcal N}_u$. On the other hand, an individual is born at each successful reproduction attempt (an unsuccessful predation attempt will occur whenever the passive is an occupied site). Assuming that the probability of the passive being an empty site is equal to $1-\rho_D$, where $\rho_D$ is equal to the density of individuals deep inside the domains, the equilibrium condition may be written as
\begin{equation}
p=r(1-\rho_{D})\,,
\end{equation}
or, equivalently, 
\begin{equation}
\rho_{D}=1-\frac{p}{r}\,. \label{rhod}
\end{equation}
According to Eq. (\ref{rhod}) equilibrium is only possible if $p<r$. If $p>r$ then $\rho_D=0$.

In practice, the probability of a site far from the borders being occupied is not independent of the distribution of individuals in neighbouring sites, and Eq. (\ref{rhod}) is only approximately valid. Fig. 8 shows the comparison between the values of $\rho_D$ obtained using the analytical approximation given by Eq. (\ref{rhod}) (dashed black line) with the corresponding numerical results (blue dots connected with a solid line) for various values of $p/r$ (assuming that $m=0.5$ and $p+r+m=1$). Fig. 8 shows that the analytical fit is almost perfect for small values of $p/r$ ($p/r < 0.5$), but that the numerical results  deviate from the analytical prediction for larger values of $p/r$. In particular, the maximum value of $p/r$ consistent with $\rho_D \neq 0$ ($p/r \sim 0.75$) is slightly smaller than the analytical prediction ($p/r = 1$).

Fig. 9 shows the probability of extinction $P_{\rm ext}$ of at least one species as a function of $p/r$ for ${\mathcal N}_u=2$ (green dots connected with a solid line) and ${\mathcal N}_u=25$ (magenta squares connected with a solid line). The results were obtained from an average of $1000$ $200^2$ simulations with two species ($N=2$) starting from random initial conditions, again assuming that $m=0.5$ and $p+r+m=1$ (we verified that the probability profiles for $N=3$ are almost identical to those obtained for $N=2$). Fig. 9 shows that the death by starvation, if $p/r$ is sufficiently low, contributes to the preservation of coexistence. The similarity of the probability profiles obtained with ${\mathcal N}_u=2$ and ${\mathcal N}_u=25$ (except for a small shift in $p/r$) indicates that the main qualitative results obtained in this paper do not depend on the specific choice of ${\mathcal N}_u$. For fixed values of $m$, $r$ and $p$, the asymptotic  characteristic length scale of the network is larger for larger values of ${\mathcal N}_u$ (note that for ${\mathcal N}_u=\infty$, the value of $L$ would always keep growing). Therefore, $L$ becomes larger for ${\mathcal N}_u=25$ than for ${\mathcal N}_u=2$. This is the reason why the interval of $p/r$ allowing for coexistence is broader for ${\mathcal N}_u=2$ than in the ${\mathcal N}_u=25$ case for simulations performed in a finite box.

\section{CONCLUSIONS \label{conclusion}}

In this letter we have investigated the dynamics of spatial stochastic May-Leonard models in two spatial dimensions, considering the death of individuals by starvation after a given number of successive unsuccessful predation attempts as a new ingredient. We have considered models with mutual predation interactions between between any two individuals of different species which, in their standard version, generally lead to the loss of coexistence in a finite time. By considering numerical simulations with two and three species, as well as analytical arguments, we have demonstrated that death by starvation can play an important role on the dynamics of population networks. In particular, we have shown that, if the reproduction rate is sufficiently high, death by starvation may lead be responsible for the preservation of coexistence.

\acknowledgments

P.P.A. acknowledges the support by FEDER—Fundo Europeu de Desenvolvimento Regional funds through the COMPETE 2020—Operational Programme for Competitiveness and Internationalisation (POCI), and by Portuguese funds through FCT - Fundação para a Ciência e a Tecnologia in the framework of the project POCI-01-0145-FEDER-031938 and the FCT grant UID/FIS/04434/2013. B.F.O thanks Funda\c c\~ao Arauc\'aria, Fapern, FCT and INCT-FCx (CNPq/FAPESP) for financial and computational support.

\end{document}